\documentclass[twocolumn,resetfootnote]{aastex701}

\graphicspath{{./}{}}
\usepackage{amsmath}
\usepackage{comment}
\usepackage{wasysym}
\usepackage{booktabs}

\begin{document}

\title{The Illusory Precision of TTV Masses: Hidden Solutions Behind Kepler-9's Tight Mass Ratio}

\correspondingauthor{Sheng Jin}
\email{jins@ahnu.edu.cn}

\author[0000-0002-9063-5987]{Sheng Jin}
\affiliation{Department of Physics, Anhui Normal University, Wuhu 241002, China}
\email{jins@ahnu.edu.cn}

\author[0000-0001-9424-3721]{Dong-Hong Wu}
\affiliation{Department of Physics, Anhui Normal University, Wuhu 241002, China}
\email{wudonghong@ahnu.edu.cn}

\author{Xiao-Ling Xu}
\affiliation{Department of Physics, Anhui Normal University, Wuhu 241002, China}
\email{3226486559@qq.com}

\author[0000-0002-9260-1537]{Jianghui Ji}
\affiliation{Purple Mountain Observatory, Chinese Academy of Sciences, Nanjing 210023, China}
\email{jijh@pmo.ac.cn}

\begin{abstract}
Transit timing variations (TTV) are considered a tool for constraining the masses of transiting planets  in the absence of radial-velocity data. 
Although theoretical studies have long revealed that TTV mass determinations intrinsically suffer from degeneracies, existing analyses of TTV data typically report a single-mode solution under a model with a specified number of planets.
This is because fitting TTV curves in the high-dimensional solution space of TTV posterior is extremely challenging; even locating a single solution requires substantial computational resources.
We developed an efficient mode-first searching algorithm that can locate multiple solutions in a single MCMC run. 
We applied this algorithm to Kepler-9 b and c, which have the highest-quality TTV data.
We found that the observed TTV can be reproduced by many combinations of planetary masses spanning a broad range, rather than the previously assumed precise determination.
The mass of Kepler-9 b can range from 31.6 to 47.1 $M_{\oplus}$, while that of Kepler-9 c can range from 21.8 to 32.3 $M_{\oplus}$, and even more broadly under looser constraints.
These degenerate solutions follow a linear relationship under a tight mass ratio between the two planets, consistent with previous theoretical predictions.
Furthermore, we demonstrate that achieving a globally converged posterior distribution for Kepler-9's TTV is impossible using a sampling algorithm that preserves the Markovian property. This underscores the need for caution when interpreting results from sampling algorithms that lack mathematical guarantees of global convergence.

\end{abstract}
\keywords{Exoplanet -- Transit timing variation}

\section{Introduction}
\label{sec:intro}

Transit timing variations (TTV), which are precise measurements of how planetary orbits deviate slightly from the ideal Keplerian trajectories expected in a single-planet system \citep{Miralda2002,Agol2005,Holman2005}, have long been regarded as a powerful tool for  determining the masses of planets in multi-planet systems \citep{Holman2010,Cochran2011,Lithwick2012,Wang2018,Wu2018,Borsato2019}, and for searching those systems for additional planets \citep{Ballard2011,Fabrycky2012,Ford2012,Steffen2012,Steffen2013,Xie2013,Yang2013,Gillon2017,Wu2023,Sun2025}. 
Given that many transiting systems lack high-precision radial velocity (RV) observations, it is only natural that we hold the TTV method in high regard, as it is the only viable method in such cases.

However, theoretical studies have demonstrated that TTV are affected by mass–eccentricity degeneracies \citep{Ballard2011,Lithwick2012,Boue2012,Schmitt2014}, and the solution space of TTV fits is known to be highly multimodal \citep{Yahalomi2024,Lammers2026}. This  is a natural expectation because TTV provide indirect measurements of planetary gravitational perturbations. They essentially encode a distinct pattern of these perturbations. Therefore, it is plausible that similar TTV can be reproduced by different combinations of planetary parameters within the solution space.

Despite theoretical studies showing that TTV solutions suffer from degeneracies, existing analyses of TTV data typically report a single-mode solution under a model with a specified number of planets. This is because locating multimodal solutions using observed TTV curves is challenging due to the intrinsic sharpness of the likelihood peaks \citep{Veras2011}. Additionally, most sampling algorithms lack the ability to locate spiky solution modes that are widely separated, which is the typical geometry of high-dimensional TTV posterior distributions.

Here, we introduce a parallel searching algorithm designed to identify as many well-fitting modes as possible across the full parameter space, explicitly accounting for the known multimodal nature of TTV mass determination. We demonstrate the effectiveness of this algorithm using the Kepler-9 system as a benchmark, which can be regarded as one of the highest-quality cases for TTV studies, as both planets have sufficiently accurate TTV measurements spanning many orbital periods \citep{Holman2010,Borsato2014,Dreizler2014,Hadden2017,Freudenthal2018,Wang2018,Borsato2019}.

The Kepler-9 system, which features two planets with an orbital period ratio close to the 2:1 mean motion resonance (MMR), was the first multi-planet system discovered using the transit method and the first to have planetary masses determined from TTV \citep{Borsato2014,Dreizler2014,Hadden2017,Freudenthal2018,Wang2018,Borsato2019}.
The latest independent TTV analyses have converged to consistent masses for Kepler-9 b and c (in units of $M_{\oplus}$): 45.1 $\pm$ 1.5 and 31.0 $\pm$ 1.0 \citep{Dreizler2014},
43.5 $\pm$ 0.6 and 29.8 $\pm$ 0.6 \citep{Borsato2014},
43.4$^{+1.6}_{-2.0}$ and 29.8$^{+1.1}_{-1.3}$ \citep{Borsato2019},
43.97 $\pm$ 0.49 and 30.24 $\pm$ 0.33 \citep{Wang2018}.
However, even for the Kepler-9 system, which is thought to have precise mass determinations based on its highest-quality TTV measurements, we identified numerous solutions spanning a large mass range.
Remarkably, these multimodal solutions exhibit a perfect linear relationship in the mass-mass space, confirming theoretical predictions that TTV in a two-planet system constrain only the mass ratio \citep{Lithwick2012}.

Additionally, we illustrate the inherent difficulties of TTV fitting using 78 million synthetic TTV curves generated from a theoretical model.
Specifically, we first simulate theoretical TTV curves for a particular planetary parameter set of the Kepler-9 system, thereby defining the true orbital parameter values. We then generate the 78 million synthetic curves by applying small perturbations around these true values and examine the resulting variation in $\chi^2$ across the parameter space.
We demonstrate that the peak of the TTV posterior distribution in the solution space is extremely sharp, consistent with the findings of \citet{Veras2011}. 
The sparse, sharp geometric of the high dimentional TTV solution space  motivated our parallel-searching algorithm: the extreme sharpness of the likelihood peak forces standard Markov chain Monte Carlo (MCMC) methods to adopt very small Gaussian proposal step sizes to efficiently explore the local vicinity of each mode. Such small step sizes, however, are insufficient to probe the full parameter space and can easily miss additional solution modes.
Such a paradox cannot be reconciled; hence, we developed a parallel searching algorithm that aims to locate the multiple modes as a priority.

This paper is organized as follows. Section \ref{sec:model} describes our parallel-searching algorithm, developed specifically for TTV fitting.
Section \ref{sec:results} presents the fitting results for the TTV of the Kepler-9 system and reveals a linear degeneracy between the planetary masses.
Section \ref{sec:space} demonstrates that the likelihood peak in TTV fitting is extremely sharp and non-Gaussian.
Section \ref{sec:discussion} provides a summary and discussion.

\section{Methods}
\label{sec:model}

\subsection{Difficulties in TTV Fitting}

The solution to a TTV curve is typically found by employing Bayesian inference, with MCMC being utilized to sample the posterior parameter space\citep{Tuchow2019}. However, constraining the planetary parameters underlying a TTV curve with MCMC methods is intrinsically challenging for two main reasons.

First, N-body simulations show that TTV amplitudes can vary by orders of magnitude when any single orbital element is perturbed by as little as $10^{-3}$ AU in semimajor axis, 0.005 in eccentricity, or a few degrees in the relevant orbital angles \citep{Veras2011}.
Consequently, the likelihood surface exhibits an extremely sharp peak in the full parameter space, forcing MCMC samplers to adopt prohibitively small step sizes (Gaussian proposals used in Monte Carlo sampling) to achieve convergence around a mode.

Second, theoretical studies have shown that TTV are subject to pronounced parameter degeneracies, making the solution space highly multimodal \citep{Ballard2011,Lithwick2012,Boue2012,Schmitt2014,Yahalomi2024,Lammers2026}.
From this perspective, an MCMC sampler must adopt a relatively large step size to ensure that other modes in the solution space are not overlooked.

Consequently, a paradox emerges: these competing demands impose contradictory sampling requirements, rendering it practically impossible for standard MCMC methods to explore the TTV parameter space globally and achieve true convergence. 
As a result, literatures showing apparently converged sampling results merely reflect adequate exploration of the vicinity of a single likelihood peak and do not guarantee comprehensive coverage of the full parameter space. In short, such solutions are locally, rather than globally, convergent.

\subsection{Mode-First Strategy}

In this work, we implement a mode-first parallel-searching algorithm whose primary objective is to identify---rather than fully characterize---each significant mode of the TTV posterior space. 
This approach is specifically designed to accommodate the degenerate, sparse, and spiky multimodal nature of TTV inversion, with the objective of identifying as many high-quality orbital solutions as possible lying behind a TTV curve.

The algorithm initiates a set of parallel MCMC chains, each of which is assigned with a distinct proposal width that determines its effective step size across the parameter space. Periodically, the chains exchange their proposal widths, ensuring that some walkers remain tuned for local refinement with small steps, while others are free to explore widely separated modes with large steps.

In practice, we launch 15 independent chains, each assigned a fixed proposal width drawn from a predefined set of ratios---(0.01, 0.005, 0.002, 0.001, 3e-4, 1e-4, 5e-5, 2e-5, 1e-5, 3e-6, 1e-6, 3e-7, 1e-7, 3e-8, 1e-8)---relative to the prior range of each parameter.
After $\sim$ 100 iterations these fixed widths are swapped among the chains.
By cyclically redistributing the proposal widths, the algorithm maintains the global-search capability of each sampler: every chain alternates between the resolution required to refine a local mode and the large stride needed to escape it. 
This enables the parallel samplers to continue migrating toward the dominant mode or any equivalently plausible global mode until no better positions are found.

To further enhance the ability of global exploration, we periodically invoke the tuning algorithm of \citet{Jin2024} (Algorithm 1) after every 5000 consecutive iterations to synthesize an expanded suite of proposal-width combinations.
An additional group of short scout MCMC chains is then launched with these newly generated widths, enabling the walkers to probe neighbouring regions and relocate to any superior likelihood peaks that are discovered.
These auxiliary proposal-width combinations are used exclusively within these short scouting phases; the main MCMC sequence---interleaved with the scouting phases---continues to rely solely on the original set of fifteen widths described above.

The principal cost of this parallel-search strategy is the loss of the Markovian property, implying that local convergence cannot be guaranteed as the proposals  are continuously changing.
We accept this trade-off because enumerating the full set of global modes is more valuable than certifying local convergence, as a sampler may have difficulty escaping the nearest peak for the TTV fitting due to the sparse and spicky features of TTV solution space. 
If locally converged posteriors are required, we can simply freeze the proposal-swapping once the samplers has settled into a mode and continue the chains with a fixed, optimally tuned proposal until standard convergence diagnostics are satisfied.

We implemented the above strategy within the \texttt{Nii-C} MCMC framework \citep{Jin2022,Jin2024}, using \texttt{TTVFast} \citep{Deck2014} to compute transit times and evaluate the likelihood.  
For the complete Kepler-9 dataset, a typical run with 15 parallel-searching chains routinely recovers 3 to 6 distinct global modes within 4,000,000 MCMC iterations. 
The corresponding orbital solutions we discovered based on the TTV curves of Kepler-9 b and c are presented in the next section.

\section{Solutions of Kepler-9 b and c} 
\label{sec:results}

\subsection{Multimodal Posterior Distribution}

We revisit the Kepler-9 system using our mode-first parallel searching strategy to determine whether the planetary masses derived from TTV are the only solutions. 
This is because, in our previous studies investigating the TTV of Kepler-9 b and c using a standard ensemble MCMC algorithm, obtaining reliable convergence results was challenging.
In some cases, we even obtained planetary masses that differed by more than their respective standard deviations across different sampling results.

\begin{table}
\caption{Prior distributions of the Kepler-9 b and c.}
\begin{tabular}{cccc}
\hline
\hline
Parameter & Prior  & Min & Max \\ 
\hline
Mass, $M_b (M_{\oplus})$ & Uniform & 10 & 50 \\ 
Period, $P_b$ (${\mathrm {days}}$) & Uniform & 19.2 & 19.4 \\ 
$e_b$ & Uniform & 0.01 & 0.1 \\ 
$i_b$ & Uniform & 88.1958 & 91.8042 \\ 
$\omega_b$ ($^{\circ}$) & Uniform & 0 & 360 \\ 
$M_{0~ b}$ ($^{\circ}$) & Uniform & 0 & 360 \\ 
Mass, $M_c (M_{\oplus})$ & Uniform & 10 & 50 \\ 
Period, $P_c$ (${\mathrm {days}}$) & Uniform & 38.9 & 39.2 \\ 
$e_c$ & Uniform & 0.01 & 0.1 \\ 
$i_c$ & Uniform & 88.8707 & 91.1293 \\ 
$\Omega_b$-$\Omega_c$ & Uniform & 0 & 20 \\ 
$\omega_c$ ($^{\circ}$) & Uniform & 0 & 360 \\ 
$M_{0~ c}$ ($^{\circ}$) & Uniform & 0 & 360 \\ 
\hline
\end{tabular}
\label{tab1}
\end{table}
\normalsize

The transit mid-times and corresponding uncertainties for Kepler-9 b and c are adopted from the TTV catalog of \citet{Holczer2016}.
We fit the TTV of Kepler-9 b and c using our mode-first parallel searching strategy with a modified version of the \texttt{Nii-C} code \citep{Jin2024}.
To fully investigate the possible solutions for the TTV of Kepler-9 b and c, we set the planetary mass ranges to a broad interval from 10 to 50  $M_{\oplus}$.
Table \ref{tab1} lists the search ranges of the 13 system parameters, which were input into \texttt{TTVFast} during each MCMC iteration to calculate the likelihood, 
assuming that the the TTV observation errors follow a Gaussian distribution.
Hence, we are using the maximum likelihood estimation (MLE) fitting on the TTV data.
Hence, we are using maximum likelihood estimation (MLE) fitting on the TTV data. We chose MLE fitting over maximum a posteriori (MAP) fitting because we did not want manually assigned priors to influence the MCMC sampling.
The orbital periods and inclinations of the Kepler-9 planets are confined to narrow ranges because these parameters are well constrained by the transit periods.
We set 4,000,000 MCMC iterations for each run. 
A typical run on an Intel Core i7-1360P laptop completes in roughly 24 hours.

\begin{table*}[htbp]
\centering
\caption{The 13 parameters for 40 solutions (from 9 MCMC runs) with reduced $\chi^2 < 10$. Values are rounded to two significant figures to save space. The appendix provides the full precision of the numbers needed to produce the exact reduced $\chi^2$.}
\label{tab:k9_param}
\begin{tabular}{@{}cccccccccccccccc@{}}
\hline
\hline
Num. &$M_b$ &$P_b$  & $e_b$ & $i_b$ & $\omega_b$ & $M_{0~ b}$ & $M_c$ & $P_c$  & $e_c$ & $i_c$ & $\Omega_b$-$\Omega_c$ & $\omega_c$ & $M_{0~ c}$ & $\chi^2$ & $M_b/M_c$ \\
 & ($M_\oplus$)& (days) &  &  &  &  & ($M_\oplus$)&  (days) &  &  &  & & &  \\
\hline
1 & 43.75 & 19.23 & 0.06 &  91.79 & 357.38 & 296.50 & 30.13 & 39.05 & 0.07 & 88.87 & 0.02 & 167.57 & 289.78 &  1.65 & 1.4481\\
2 & 39.77 & 19.23 & 0.07 & 91.19 &  359.74 & 293.37 & 27.41 &  39.04 & 0.07 & 90.86 & 1.08 & 168.52 & 288.94 & 2.13 & 1.4555 \\
3 & 39.24 & 19.23 & 0.08 & 89.51 & 2.40 & 290.20 & 27.02 & 39.04 & 0.07 & 90.88 & 6.68 & 168.94 & 289.11 & 2.27 & 1.4569 \\
4 & 39.81 & 19.23 & 0.07 & 91.78 & 0.02 & 293.01 & 27.44 & 39.04 & 0.07 & 88.90 & 0.07 & 167.73 & 289.57 & 2.39 & 1.4522 \\
5 & 38.92 & 19.23 & 0.07 & 91.80 & 0.45 & 292.34 & 26.82 & 39.04 & 0.07 & 88.89 & 0.23 & 167.96 & 289.32 & 2.90 & 1.4501 \\
6 & 40.61 & 19.23 & 0.07 & 91.76 & 0.01 & 293.22 & 27.97 & 39.04 & 0.07 & 88.87 & 0.02 & 167.09 & 290.22 & 2.98 & 1.4522 \\
7 & 43.96 & 19.23 & 0.07 & 89.60 & 0.39 & 293.16 & 30.20 & 39.05 & 0.08 & 89.73 & 8.10 & 171.33 & 287.19 & 3.21 & 1.4506 \\
8 & 41.41 & 19.23 & 0.07 & 91.08 & 0.01 & 293.40 & 28.52 & 39.04 & 0.07 & 89.00 & 3.68 & 169.47 & 288.17 & 3.28 & 1.4536 \\
9 & 44.54 & 19.23 & 0.07 & 91.71 & 357.71 & 296.09 & 30.64 & 39.05 & 0.07 & 88.87 & 6.56 & 174.22 & 283.97 & 3.28 & 1.4484 \\
10 & 41.75 & 19.23 & 0.07 & 91.65 & 0.01 & 293.32 & 28.76 & 39.04 & 0.07 & 88.90 & 5.35 & 171.70 & 286.14 & 3.31 & 1.4587 \\
11 & 44.80 & 19.23 & 0.07 & 88.79 & 0.00 & 293.81 & 30.76 & 39.05 & 0.07 & 90.04 & 7.07 & 168.60 & 289.64 & 3.36 & 1.4524 \\
12 & 41.54 & 19.23 & 0.07 & 91.17 & 0.00 & 293.40 & 28.62 & 39.04 & 0.07 & 89.10 & 3.98 & 169.61 & 288.05 & 3.40 & 1.4493\\
13 & 37.54 & 19.23 & 0.08 & 91.80 & 1.27 & 291.18 & 25.90 & 39.04 & 0.07 & 88.92 & 0.48 & 168.15 & 289.10 & 3.52 & 1.4513 \\
14 & 41.60 & 19.23 & 0.07 & 91.00 & 0.00 & 293.42 & 28.66 & 39.04 & 0.07 & 88.90 & 3.85 & 169.44 & 288.21 & 3.55 & 1.4557\\
15 & 38.62 & 19.23 & 0.08 & 90.99 & 4.65 & 287.33 & 26.57 & 39.05 & 0.08 & 90.66 & 10.11 & 173.31 & 285.58 & 4.07 & 1.4493\\
16 & 36.65 & 19.23 & 0.08 & 91.77 & 1.98 & 290.21 & 25.29 & 39.04 & 0.07 & 88.88 & 1.61 & 168.73 & 288.52 & 4.15 & 1.4490 \\
17 & 43.98 & 19.23 & 0.07 & 88.20 & 1.79 & 291.66 & 30.19 & 39.05 & 0.08 & 90.46 & 8.34 & 167.68 & 290.75 & 4.31 &1.4480 \\
18 & 44.97 & 19.23 & 0.06 & 89.55 & 0.00 & 293.86 & 30.90 & 39.05 & 0.07 & 90.05 & 7.28 & 169.49 & 288.85 & 4.42 & 1.4483\\
19 & 35.17 & 19.23 & 0.09 & 90.60 & 6.76 & 284.15 & 24.22 & 39.04 & 0.08 & 90.32 & 10.10 & 172.93 & 285.74 & 4.44 &1.4556 \\
20 & 36.13 & 19.23 & 0.08 & 91.80 & 2.68 & 289.24 & 24.93 & 39.04 & 0.07 & 88.87 & 4.04 & 170.81 & 286.65 & 4.57 & 1.4502\\
21 & 35.31 & 19.23 & 0.09 & 91.46 & 4.79 & 286.61 & 24.35 & 39.04 & 0.07 & 89.05 & 7.25 & 173.31 & 284.70 & 4.92 & 1.4491\\
22 & 36.05 & 19.23 & 0.09 & 91.51 & 5.76 & 285.46 & 24.84 & 39.04 & 0.08 & 89.07 & 9.86 & 176.27 & 282.41 & 5.46 & 1.4487 \\
23 & 35.16 & 19.23 & 0.08 & 91.79 & 2.90 & 288.86 & 24.26 & 39.04 & 0.06 & 88.93 & 0.10 & 167.63 & 289.50 & 5.49 & 1.4564\\
24 & 34.27 & 19.23 & 0.09 & 91.53 & 6.54 & 284.15 & 23.63 & 39.04 & 0.07 & 89.04 & 9.46 & 175.78 & 282.68 & 5.87 & 1.4510\\
25 & 36.26 & 19.23 & 0.09 & 90.85 & 6.82 & 284.19 & 24.96 & 39.04 & 0.08 & 89.12 & 11.29 & 176.56 & 282.51 & 5.97 &1.4499 \\
26 & 34.65 & 19.23 & 0.09 & 91.52 & 6.76 & 283.92 & 23.89 & 39.04 & 0.08 & 89.05 & 10.20 & 176.55 & 282.10 & 6.13 &1.4484 \\
27 & 36.19 & 19.23 & 0.09 & 91.31 & 6.58 & 284.44 & 24.92 & 39.04 & 0.08 & 89.08 & 11.11 & 177.13 & 281.87 & 6.26 &1.4523 \\
28 & 35.27 & 19.23 & 0.09 & 91.46 & 7.19 & 283.51 & 24.31 & 39.04 & 0.08 & 89.04 & 11.11 & 177.23 & 281.71 & 6.48 &1.4512 \\
29 & 45.78 & 19.23 & 0.06 & 89.65 & 0.00 & 294.04 & 31.43 & 39.05 & 0.07 & 91.13 & 7.00 & 167.39 & 290.85 & 6.73 &1.4485 \\
30 & 33.03 & 19.23 & 0.09 & 91.45 & 6.13 & 284.50 & 22.81 & 39.04 & 0.07 & 88.92 & 7.08 & 172.85 & 284.93 & 6.83 & 1.4516\\
31 & 33.47 & 19.23 & 0.09 & 91.73 & 4.16 & 287.03 & 23.11 & 39.04 & 0.06 & 89.03 & 1.74 & 168.12 & 288.97 & 7.18 &1.4522 \\
32 & 33.09 & 19.23 & 0.09 & 88.20 & 5.33 & 285.60 & 22.85 & 39.04 & 0.07 & 91.13 & 3.78 & 165.36 & 291.78 & 7.47 & 1.4518\\
33 & 32.82 & 19.23 & 0.09 & 91.78 & 4.62 & 286.30 & 22.67 & 39.04 & 0.06 & 88.95 & 2.85 & 169.13 & 288.00 & 7.86 & 1.4491\\
34 & 39.33 & 19.23 & 0.09 & 88.59 & 7.29 & 284.37 & 27.00 & 39.04 & 0.08 & 90.88 & 12.21 & 169.35 & 289.82 & 8.17 & 1.4563\\
35 & 47.11 & 19.23 & 0.06 & 89.44 & 0.00 & 294.14 & 32.30 & 39.05 & 0.08 & 90.21 & 8.62 & 169.81 & 288.88 & 8.19 & 1.4570\\
36 & 32.31 & 19.23 & 0.09 & 88.20 & 6.41 & 284.15 & 22.30 & 39.04 & 0.07 & 91.08 & 4.93 & 165.07 & 292.16 & 8.23 & 1.4511 \\
37 & 42.23 & 19.23 & 0.08 & 90.99 & 3.64 & 288.99 & 29.01 & 39.05 & 0.08 & 89.44 & 11.49 & 176.30 & 283.05 & 8.39 &1.4518 \\
38 & 32.10 & 19.23 & 0.10 & 88.21 & 6.52 & 283.95 & 22.16 & 39.04 & 0.07 & 91.13 & 4.97 & 165.07 & 292.16 & 8.51 & 1.4522\\
39 & 32.18 & 19.23 & 0.09 & 88.20 & 6.30 & 284.25 & 22.22 & 39.04 & 0.07 & 91.12 & 4.53 & 164.94 & 292.21 & 8.58 & 1.4510\\
40 & 31.58 & 19.23 & 0.10 & 88.20 & 6.68 & 283.58 & 21.80 & 39.04 & 0.07 & 91.11 & 5.06 & 165.11 & 292.06 & 9.09 & 1.4538\\
\hline
\end{tabular}%
\label{tab2}
\end{table*}

\begin{figure*}
\centering
\includegraphics[width=1.0\textwidth]{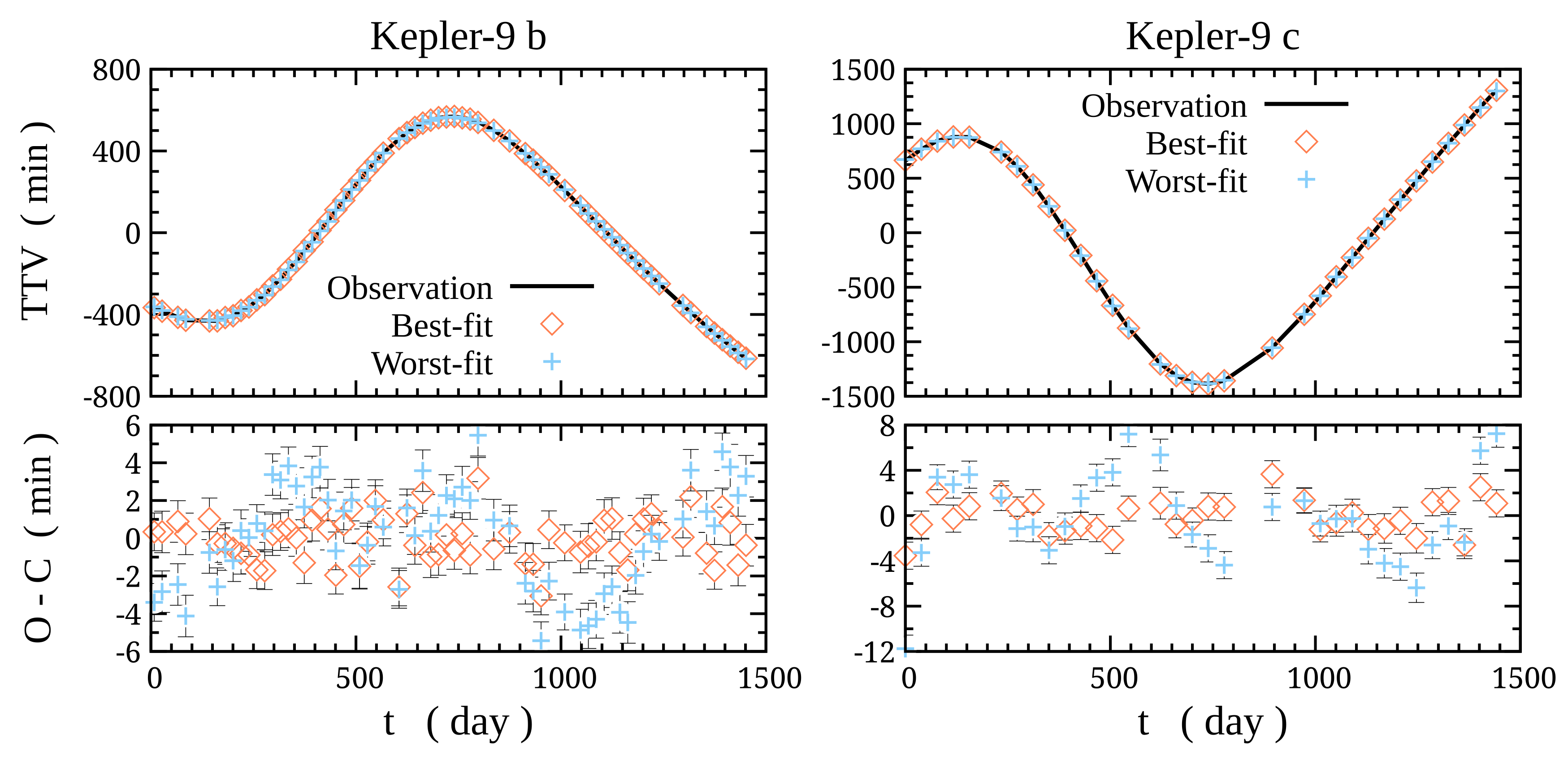}
\caption{The TTV and residual curves for the first (best-fit) and 40th (worst-fit) solutions in Table 2. Both reproduce the TTV of Kepler-9 b and c well; only the 40th solution yields larger residuals.}
\label{fig:1}
\end{figure*}

The mode-first searching strategy proves highly efficient in addressing the Kepler-9 TTV  and immediately reveals a striking feature: the posterior distribution is strongly multimodal, with several distinct mass combinations that fit the TTV of Kepler-9 b and c equally well.
Typically, a mode-first searching MCMC with 15 parallel workers can locate approximately 3 to 6 combinations of planetary parameters that accurately explains the TTV of Kepler-9 system.
Furthermore, the multimodal solutions in a run exhibit significant differences in planetary masses.

To investigate the multimodal features underlying the TTV curves of Kepler-9 b and c, we conducted 9 independent mode-first searching MCMC runs  with identical initial configurations, except for different initial random seeds.
It turns out that the more independent mode-first searching MCMC runs we perform, the more distinct modes we can identify.
Table \ref{tab2} lists the 40 solutions with the lowest reduced $\chi^2$ values.
All these 40  parameter combinations reproduce the observed TTV satisfactorily, each yielding reduced $\chi^2$ $<$ 10.
The 40 best-fitting solutions uncovered by 9 independent MCMC runs span a broad mass range: Kepler-9 b, its mass ranges from 31.58 to 47.11 $M_{\oplus}$ and Kepler-9 c  from 21.80  to 32.30  $M_{\oplus}$, corresponding to $> 50$ percent variation for each planet.

Although the upper entries in Table \ref{tab2}  yield slightly lower reduced  $\chi^2$ 
 values, all 40 solutions reproduce the observed TTV satisfactorily. 
 Figure \ref{fig:1} compares the observed TTV curves with the theoretical curves corresponding to the first and 40th entries (the best-fit and the worst-fit cases):  both track the observations closely, with only marginally larger residuals for the 40th-ranked case.
Note that the higher-ranked  entries in Table \ref{tab2} can deliver noticeably smaller residuals than the 40th-ranked case.
 These results demonstrate that the posterior distribution of the Kepler-9 system's TTV is indeed multimodal.

\subsection{Linear Mass Degeneracy}

\begin{figure*}
\centering
\includegraphics[width=1.0\textwidth]{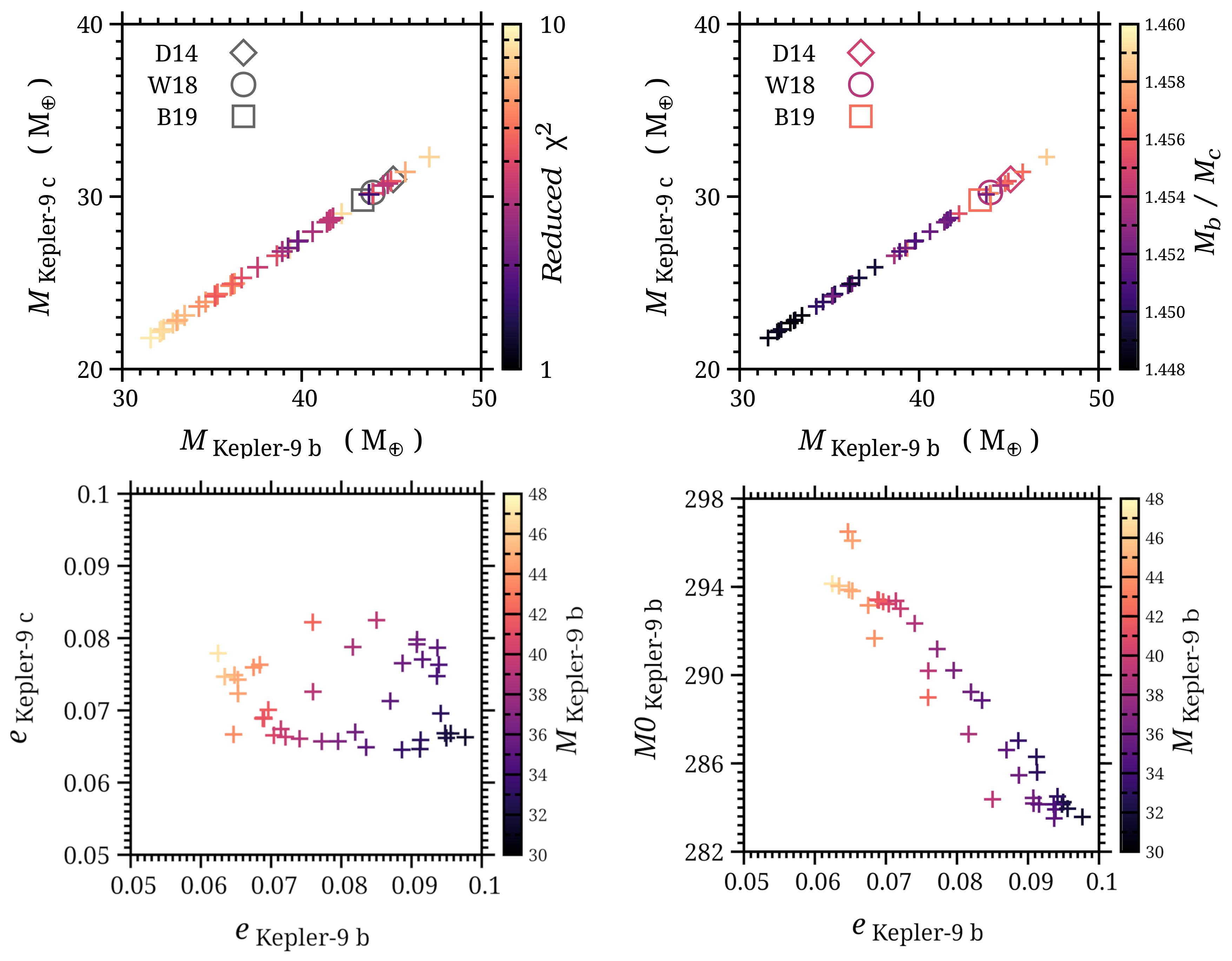}
\caption{
The top two panels show the masses of Kepler-9 b and c in an $M_b$ versus $M_c$ plane, revealing a perfect linear correlation.
In the top left panel, points are color‑coded by the reduced $\chi^2$ between the observed TTV curve and the theoretical curve for each of the 40 solutions. The top right panel instead colors points by the corresponding mass ratio. For comparison, masses derived  from previous studies are also plotted and labeled as D14 \citep{Dreizler2014}, W18 \citep{Wang2018}, and B19 \citep{Borsato2019}.
D14 used a stellar mass of $1.05\pm0.03$ $M_{\odot}$, B19 used $1.07\pm{0.05}$ $M_{\odot}$, W18 and this work used $1.034$ $M_{\odot}$---such small differences do not significantly affect the derived planetary masses.
The bottom two panels present two additional parameter relationships identified across the 40 solutions. The bottom left panel shows a linear (though not strict) anti-correlation between the eccentricities of the two planets: solutions with lower total masses tend to favor larger eccentricities for Kepler-9 b and smaller eccentricities for Kepler-9 c. The bottom right panel reveals a linear trend between the mean anomaly $M_0$ and eccentricity of Kepler-9 b, such that lower‑mass solutions correspond to a larger eccentricity and a smaller $M_0$ for Kepler‑9 b.
}
\label{fig:2}
\end{figure*}

The wide range in the planetary masses of Kepler-9 b and c, as listed in Table 2, suggests that accurate mass constraints cannot be obtained based on the TTV. However, the masses of Kepler-9 b and c exhibit an underlying relationship despite their widely distributed values: the mass ratio between Kepler-9 b and c remains consistent. All 40 solutions show a mass ratio of approximately 1.45, with values ranging from 1.4481 to 1.4569; the extreme values differ by less than one percent.

The top two panels of Figure \ref{fig:2} show the two-planet mass distribution in the  Kepler-9 b versus  Kepler-9 c mass plane ($M_b-M_c$), revealing the expected tight linear correlation arising from their similar mass ratio.
Along this ridge the lowest reduced $\chi^2$ values cluster in two distinct regions: one  at $M_b$ $\sim$ 40 $M_{\oplus}$ and  $M_c$ $\sim$  27 $M_{\oplus}$ and the other one  at $M_b$ $\sim$ 44 $M_{\oplus}$ and  $M_c$ $\sim$  30 $M_{\oplus}$.
Both regions yield  a reduced $\chi^2$ $\sim$ 2, indicating comparably excellent fits to the observed TTV data.

The top right panel of Figure \ref{fig:2} further reveals that, although the planetary mass ratio is tightly constrained, a weak trend persists: lower-mass pairs exhibit slightly smaller ratios than their higher-mass counterparts.
Figure \ref{fig:2} also shows the derived planetary masses from previous research \citep{Dreizler2014,Wang2018,Borsato2019}, all of which lie along the linear ridge\footnote{
Note that the stellar masses adopted in these studies differ slightly: \citet{Dreizler2014} used $1.05\pm0.03$ $M_{\odot}$, \citet{Wang2018} and this work used $1.034$ $M_{\odot}$, while \citet{Borsato2019} used $1.07\pm{0.05}$. This may lead to variations in the derived planetary masses.
}.

Although some solutions in Figure \ref{fig:2} yield larger reduced $\chi^2$ values, they still provide acceptable fits to the observed TTV.
As Figure \ref{fig:1} shows, the worst point in the worst-case curves deviates from the observation by only about 10 minutes, a gap comparable to typical  TTV measurement errors or to any extra TTV signal that might be introduced by the candidate low-mass planet Kepler-9 d \citep{Torres2011}.
Therefore, even the poorest-fitting model cannot be rejected outright.
Furthermore, these 40 solutions stem from just nine independent mode-first MCMC searches.
Additional MCMC runs would undoubtedly populate the linear ridge more densely and could  reveal equally good fits elsewhere along it.

The tight constraint on the mass ratio can be explained theoretically. 
The amplitudes of the complex TTV of a planet can be approximated by a combined form in which the variables consist only of the perturbing planet's mass and the complex conjugate $Z_{\rm free}$, where $Z_{\rm free}$ a linear combination of the free complex eccentricities of the two planets \citep{Lithwick2012}.
Hence, in cases where both planets' periods have been observationally constrained, as in the Kepler-9 system, one can determine their mass ratio from the TTV amplitudes, but not their individual masses without knowing $Z_{\rm free}$ \citep{Lithwick2012}.

To identify the origin of the linear mass relationship between the two planets, we searched for additional degeneracy trends among the 40 solutions, examining each pair of orbital parameters for both planets.
As a result, we identify two additional distinct relationships between two pairs of orbital parameters, as presented in the bottom two panels of Figure \ref{fig:2}.
The first is a linear---though not strict---correlation between the eccentricities of Kepler-9 b and c, meaning that for solutions with lower planetary masses, Kepler-9 b generally has a larger eccentricity while Kepler-9 c has a smaller one.
The second is a linear trend between the mean anomaly $M_0$ and the eccentricity of Kepler-9 b, meaning that for solutions with lower planetary masses, Kepler-9 b generally exhibits a larger eccentricity and a smaller $M_0$.
These trends provide further insight into the linear mass degeneracy behind the TTV curves of Kepler-9 b and c, suggesting that the two TTV curves  can be well reproduced by adjusting the planetary masses, eccentricities, and mean anomalies. Specifically, the planetary masses may span a broad range, accompanied by only limited variations in eccentricities and minimal changes in $M_0$.

TTV have long been regarded as a powerful means of inferring the masses of transiting planets for which RV follow-up is unavailable.
However, our investigation of the TTV of Kepler-9 b and c shows that the TTV signal almost perfectly constrains the mass ratio while leaving the individual masses free to slide along a linear ridge.
Previous studies have all derived the masses of Kepler-9 b and c around one mode due to the spiky geometry of the TTV solutions in high-dimensional space. However, our results show that the mass ratio degeneracy allows the planetary masses to span a very large range. This leaves the individual masses of the planets in the Kepler-9 system undetermined once again.

It is also important to note that the Kepler-9 system represents an ideal case for TTV observations, where TTV for both planets have been obtained; however, severe mass degeneracies still exist.
In less ideal cases, such as when only one planet's TTV can be observed, there are even more degeneracies, as even the mass ratio can vary largely.

\section{Morphology of a theoretical mode}
\label{sec:space}

\begin{figure*}
\centering
\includegraphics[width=1.0\textwidth]{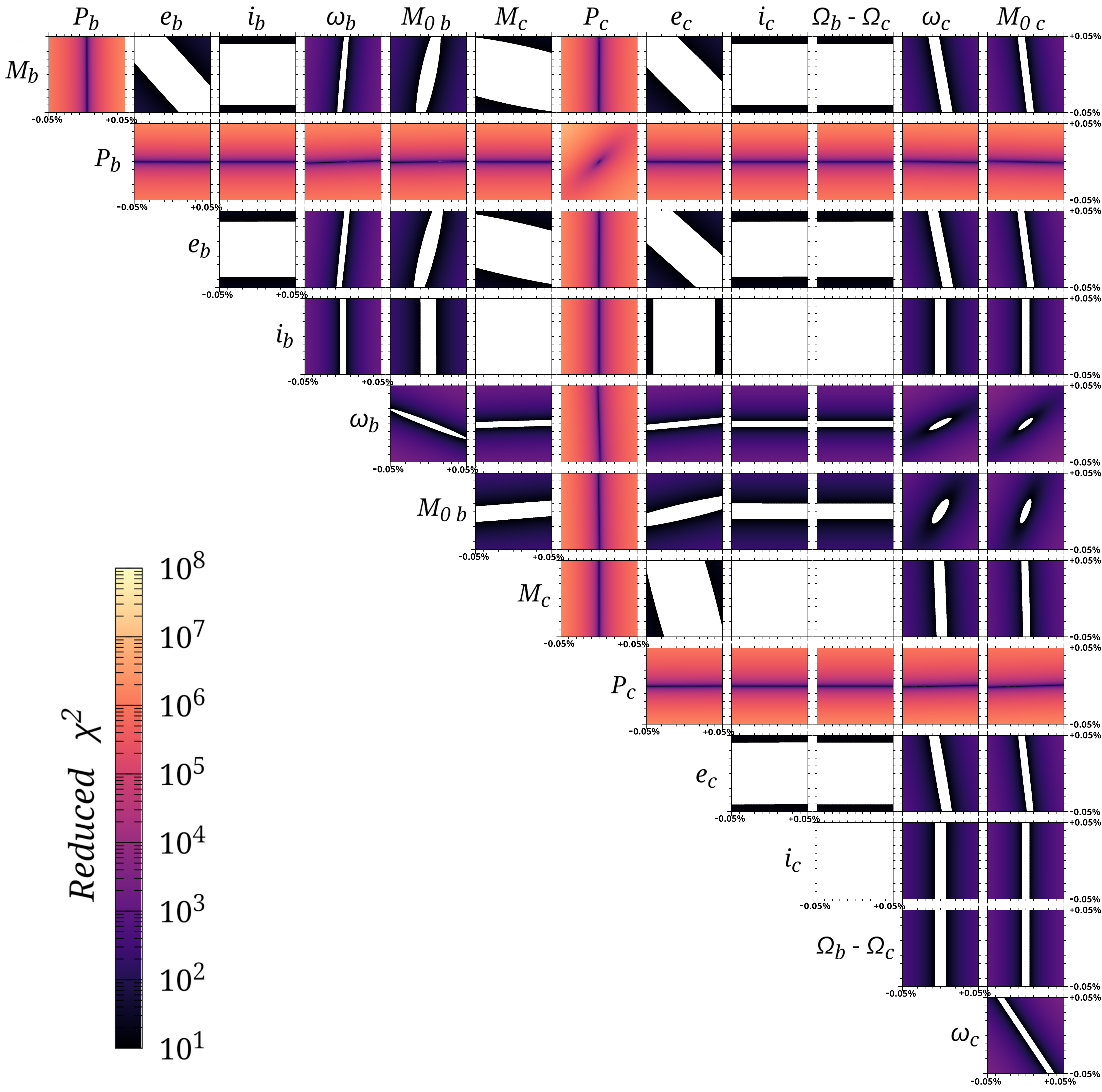}
\caption{A manually constructed corner plot quantifying  the variation of the reduced  $\chi^2$ around the true parameters of a synthetic TTV data.
White regions indicate areas where the reduced  $\chi^2 < 10$.
Each panel shows the joint behavior of one parameter pair, with the remaining 11 parameters fixed at their true values. 
Each parameter is varied by $\pm0.05\%$ around its ture values.
The strongly non-elliptical contours demonstrate that most of the probability density in the solution space for the parameters pairs is non-Gaussian. 
}
\label{fig:3}
\end{figure*}

The multimodal posterior distribution that we uncovered differs from the single-peaked, nearly Gaussian distributions that are usually reported for TTV fits of Kepler-9 system---distributions that are usually taken as proof that an MCMC chain has converged \citep{Borsato2019}.
These apparently converged Gaussians typically exhibit narrow standard deviations for planetary masses, and the narrow Gaussian-shaped contours are then used to suggest global convergence of the sampling and well-constrained planetary masses.
However, our results show that in the linear ridge of mass degeneracy reported in this work, the planetary masses of the Kepler-9 b and c can differ by 50\%.
A straightforward explanation for this discrepancy is that the solution space for TTV fitting is inherently multimodal; the Gaussian-shaped posterior distributions reported in earlier studies simply sample one of the multiple modes, and thus do not achieve global convergence.

To investigate the intrinsic geometry of the TTV solution space, we generated synthetic TTV  curves for Kepler-9 b and c using manually set planetary parameters,  assigning each synthetic TTV point a measurement uncertainty of one minute.
The planetary masses are $M_b$ = 43.98 $M_{\oplus}$ and $M_c$ =  30.25 $M_{\oplus}$, and the orbital periods are $P_{b}$ = 19.226 days and  $P_c$ = 30.015 days.
We used synthetic TTV curves because, in this case, we know the true values behind them. 
In contrast, when analyzing the observed TTV of the Kepler-9 system, the true values of the system parameters are unknown.
Consequently, when the theoretical TTV curves computed using these true values are compared with the synthetic observations, the reduced $\chi^2$ is precisely 0.
These true values define the mode of the TTV likelihood in the full 13-dimensional parameter space.
Then, we perturb each parameter pair in turn---while holding the others fixed at their true values---and record how the reduced $\chi^2$ changes with tiny deviations.
In other words, we map the local geometry around the mode via a sequence of two-dimensional slices around the mode.

In practice, we vary each parameter pair over a total range of 0.1\% ($\pm.05$\% for each individual parameter). We construct a uniform grid of 1000 points per parameter, resulting in 1,000,000 total TTV evaluations to investigate the response surface across the 0.1\% $\times$ 0.1\% two-parameter deviation space.
This results in a total of 78,000,000 TTV calculations for all the parameter pairs.
Figure \ref{fig:3} presents the corresponding manually constructed corner plot that quantifies the variation of the reduced  $\chi^2$ in the vicinity of the true parameters used to generate our synthetic TTV data.

The most impression in Figure \ref{fig:3} is that the TTV fitting landscape in the solution space features a very steep peak, especially for the orbital periods: a 0.05\% shift in either planet's orbital period can increase the reduced $\chi^2$ from 0 to on the order of 1 million.
This aligns with the results of \citep{Veras2011}, which demonstrate that TTV signal amplitudes may vary by orders of magnitude due to slight variations in any one orbital parameter.
This explains why previous studies have all constrained the masses of Kepler-9 b and c with tiny standard deviations \citep{Borsato2014,Dreizler2014,Wang2018,Borsato2019}, as the posteriors increase sharply away from the theoretical true values of TTV data.
Consequently, to sample this peak using an MCMC algorithm that preserves Markovian  properties (e.g., by adopting a constant proposal size), we must employ tiny proposal sizes to ensure a sufficient acceptance rate, allowing us to accumulate enough accepted points near the posterior peak and thoroughly explore the region around it.
However, such tiny proposals cannot effectively explore the global parameter space. 
This paradox highlights the practical challenges of TTV fitting. It also explains why, when a mode-first approach is used, we identify 40 solutions spanning a wide mass range, whereas previous studies located only one solution.

The second impression is that most two-dimensional slices through the TTV solution space  are far from Gaussian; instead, they exhibits strong linear correlations.
Consequently, a properly converged TTV posterior will not resemble a symmetric ellipse, and a corner plot that appears Gaussian does not indicate convergence. 
MCMC algorithms that exhibit collective behaviors among many walkers tend to aggregate in the space around the dominant modes; thus, we need to be cautious when interpreting results from such sampling algorithms.

\section{SUMMARY and DISCUSSION}

\label{sec:discussion}

The major finding of our work is that, even for the Kepler-9 system---which has the highest-quality TTV data for both planets---the planetary masses behind the TTV curves  remain highly multimodal.
There are many equally acceptable solutions lie on a linear ridges of nearly fixed mass ratios, with planetary masses can vary by 50\% while preserving the observed TTV.
The tight linear mass-ratio constraint and the subtle increasing trend of the mass ratio along the 40 acceptable solutions shown in Figure \ref{fig:2} confirm theoretical studies that degeneracy exists under a mass ratio determined by two-planets' TTV curves \citep{Lithwick2012}.
Our results suggest that previous mass estimates for the planets in the Kepler-9 system reflect sampling from only one mode in the full solution space \citep{Borsato2014,Dreizler2014,Wang2018,Borsato2019}.

Note that such mass degeneracy in the Kepler-9 system can be broken by high-presission RV observations.
In the discovery paper, \citet{Holman2010} included six RV measurements from Keck-HIRES, obtaining RV semi-amplitudes of $\simeq$ 19 and $\simeq$ 10 (m s$^{-1}$) for Kepler-9 b and c, respectively.
Combining these with 7 months of TTV data, they derived planetary masses of $80.0\pm4.1$ and $54.3\pm4.1$ $M_{\oplus}$ for Kepler-9 b and c.
Subsequently, \citet{Borsato2014} collected 30 RV observations with HARPS-N, which yielded lower semi-amplitudes of $\simeq$ 10 and $\simeq$ 6 (m s$^{-1}$) for Kepler-9 b and c.
Using these data, \citet{Borsato2019} performed a joint RV and TTV fit, which significantly reduced the planetary masses to $43\pm2$ and $30\pm1$ $M_{\oplus}$ for Kepler-9 b and c, respectively.
In the most recent RV observations, \citet{Weiss2024} collected 21 RV measurementss with Keck-HIRES spanning 2010 to 2022.
They derived RV semi-amplitudes of $6.1\pm2.3$ and $5.1\pm2.7$ (m s$^{-1}$) for Kepler-9 b and c, respectively, correspoinding to mass constraints of $26\pm10$ and $27\pm14$ $M_{\oplus}$ based solely on the RV data \citep{Weiss2024}.
The planetary masses from our 40 solutions are consistent with the range allowed by the latest RV observations. 
However, we did not perform a joint RV and TTV fit in this work, as the primary goal of this paper is to investigate the degeneracy inherent in TTV data alone.
Future high-precision RV observations can provide tighter constains on the Kepler-9 system; such an analysis is beyond the scope of the present study.

As the first discovered two-planet system, Kepler-9 b and c have attracted significant attention because their TTV curves have been used to constrain their masses.
The two planets are also in a 2:1 MMR, so the gravitational perturbations between them are stronger than in non-resonant cases. 
Many multi-planet systems that display measurable TTV are also locked in MMRs \citep{Xie2013,Deck2016,Nesvorny2016,Agol2021}, and both \citet{Xie2013} and \citet{Nesvorny2016} have demonstrated that the TTV amplitude directly encodes the planetary mass ratio. 
Our work extends this picture by showing that, even when the mass ratio is pinned down to percent-level precision, a significant degeneracy among the absolute mass combinations can persist. 
While the present analysis is restricted to Kepler-9, we will expand the survey to additional resonant and non-resonant systems in future studies.

Recently, \citet{Lammers2026} demonstrated the multimodal nature of TTV fitting from another perspective.
They performed a systematic reassessment of all 12 published cases in which a non-transiting planet was claimed to have been uniquely characterized using TTVs. 
For most of these single-planet transiting systems, they discovered  multiple viable solutions for the non-transiting perturbing planet, often involving very different combinations of planetary masses at different MMRs.
Our work broadens the picture of TTV degeneracy to the two-planet, both-transiting case,  which represents the highest-quality TTV data expected.
Although the MMR and mass ratio between the two planets are well determined in this case, multimodal solutions still exist.
These solutions are obtained by varying the masses of the two planets over a wide range while maintaining a fix mass ratio.
Moreover, the fixed mass ratio in the Kepler-9 system arises from the fact that both planets are transiting.
If only one of the Kepler-9 planets were transiting, not only would the MMR become undetermined, but the mass ratio itself would admit multiple solutions.
In fact, we carried out several mode-first MCMC fitting on the TTV curve of Kepler-9 b alone, mimicking the case of one transiting planet perturbed by a non-transiting companion.
We find that Kepler-9 b's TTVs can be explained by far more diverse solutions across a wide range of mass ratios for the non-transiting planet c, even when restricted to the same 2:1 MMR.

The intrinsic multimodal likelihood of TTV fitting poses challenges in obtaining globally converged sampling results when using MCMC algorithm to sample the Bayesian posterior.
These challenges stem from the paradox between effectively searching the global parameter space using large proposals and effectively exploring the vicinity of a mode using tiny proposals, since the peaks in the solution space of TTV fitting are very steep.
Our mode-first searching algorithm uses a dynamic cadence with a fixed combination of proposal sizes; however, such varying proposals cannot maintain the Markovian property and hence will not achieve global convergence.
In fact, due to the highly spiky geometry of the modes in the high-dimensional parameter space of TTV fitting, globally converged sampling results using fixed proposals are impossible.
Moreover, our theoretical calculations in Figure \ref{fig:3} show that most two-dimensional slices through the TTV solution space are far from Gaussian.
Hence, the symmetric, elliptical contours displayed in the corner plots of the TTV fitting for the Kepler-9 system---and perhaps in the fittings of other systems---merely explore the vicinity of one mode.
Even assuming that the MCMC algorithm used in their code preserves the Markovian property---though ensemble samplers typically do not---their results may only be considered converged within one of the many modes.

Our findings underscore the need for caution when inferring planetary masses from TTV data, as significant degeneracies exist even in ideal cases where sufficient TTV data covering many orbital periods for both planets have been obtained.
In less ideal cases---such as when limited TTV data is available, or only one planet is transiting and showing TTV---the degeneracies will be even more severe.
In future work, we will extend this analysis to a larger sample of exoplanet systems and further explore the intrinsic geometry of the TTV solution space.

\vspace{0.9cm}

We thank the anonymous referee for the constructive comments that highly improved the manuscript.
The authors warmly thank Songhu Wang, Fei Yan, Subo Dong, Darin Ragozzine, and Gabriel-Dominique Marleau for helpful discussions.
S.J. acknowledges  the financially supported by the National Natural Science Foundation of China (grant No. 11973094) and the incubation program for recruited talents from Anhui Normal University (2023GFXK153). D.W. acknowledges the supported by National Natural Science Foundation of China (Grant No. 12573076). J.J. is financially supported by the National Natural Science Foundation of China (grant No. 12533011, 12033010).

\software{\texttt{Nii-C} \citep{Jin2024}, \texttt{TTVFast} \citep{Deck2014}}



\appendix

\section{Full precision values of the 40 solutions}

\begin{table}[htbp]
\centering
\rotatebox{90}{%
\begin{minipage}{0.95\textheight}
  \def\arraystretch{1.2}
   \small\centering
   \setlength{\tabcolsep}{1.0mm}
    \scriptsize
    \centering
\caption{The full precision parameter values for the first half of the 40 solutions in Table \ref{tab2}.}
\label{tab:k9_param}
\begin{tabular}{@{}cccccccc@{}}
\hline
Num. &$M_b$ ($M_\oplus$) &$P_b$ (days) & $e_b$ & $i_b$ & $\omega_b$ & $M_{0~ b}$ & $M_c$ ($M_\oplus$) \\
& $P_c$ (days) & $e_c$ & $i_c$ & $\Omega_b$-$\Omega_c$ & $\omega_c$ & $M_{0~ c}$ & $\chi^2$  \\
\hline
1 & 4.374931660197e+01 & 1.922893613004e+01 & 6.465901498877e-02 & 9.178694662322e+01 & 3.573824409258e+02 & 2.965002698527e+02 & 3.012720942457e+01  \\
& 3.904685525566e+01 & 6.666899472189e-02 & 8.887430125094e+01 & 2.194758260275e-02 & 1.675712032205e+02 & 2.897847165904e+02 & 1.65  \\
\hline 
2 & 3.976679802525e+01 & 1.922908963858e+01 & 7.139842695291e-02 & 9.119177592267e+01 & 3.597404877655e+02 & 2.933654133728e+02 & 2.740721397078e+01  \\
& 3.904473914754e+01 & 6.741230621978e-02 & 9.085783449470e+01 & 1.080891458407e+00 & 1.685175850529e+02 & 2.889439140468e+02 & 2.13  \\
\hline
3 & 3.924050612677e+01 & 1.922906889920e+01 & 7.596926720754e-02 & 8.951244796261e+01 & 2.400646909841e+00 & 2.901951339126e+02 & 2.702124946824e+01  \\
& 3.904468891456e+01 & 7.258175228280e-02 & 9.087848181796e+01 & 6.678759117058e+00 & 1.689434782460e+02 & 2.891119564915e+02 & 2.27    \\
\hline
4 & 3.980796794877e+01 & 1.922923486126e+01 & 7.204811780365e-02 & 9.178269165601e+01 & 2.317771722783e-02 & 2.930095896629e+02 & 2.744268913677e+01  \\
& 3.904399885879e+01 & 6.631965472246e-02 & 8.889744309772e+01 & 7.191065478546e-02 & 1.677297771141e+02 & 2.895749879956e+02 & 2.39   \\
\hline
5 & 3.891666576803e+01 & 1.922925943383e+01 & 7.405707310326e-02 & 9.179543525289e+01 & 4.505588735076e-01 & 2.923412020175e+02 & 2.681895262650e+01  \\
& 3.904354738983e+01 & 6.607219844374e-02 & 8.888528638538e+01 & 2.260446921645e-01 & 1.679645192255e+02 & 2.893162299389e+02 & 2.90    \\
\hline
6 & 4.060675862795e+01 & 1.922932024923e+01 & 7.039909759559e-02 & 9.176310385569e+01 & 1.212867664089e-02 & 2.932233428211e+02 & 2.796959493808e+01  \\
& 3.904400543509e+01 & 6.654586271278e-02 & 8.887298138813e+01 & 2.429166816400e-02 & 1.670863445577e+02 & 2.902186123017e+02 & 2.98    \\
\hline
7 & 4.395624108772e+01 & 1.922868398793e+01 & 6.750926393580e-02 & 8.960461884178e+01 & 3.907228956131e-01 & 2.931615402040e+02 & 3.019578171204e+01  \\
& 3.904826656620e+01 & 7.595354227186e-02 & 8.973211722641e+01 & 8.104492550398e+00 & 1.713324851975e+02 & 2.871899567089e+02 & 3.21    \\
\hline
8& 4.454546740245e+01 & 1.922865846436e+01 & 6.529196650814e-02 & 9.170520755460e+01 & 3.577143188949e+02 & 2.960921757693e+02 & 3.064108494807e+01  \\
& 3.904844405631e+01 & 7.232425442789e-02 & 8.887341285769e+01 & 6.561863370039e+00 & 1.742159181416e+02 & 2.839680939096e+02 & 3.28    \\
\hline
9& 4.141183908732e+01 & 1.922923097282e+01 & 6.899302937852e-02 & 9.107626965468e+01 & 8.603949963863e-03 & 2.933951039108e+02 & 2.851642385638e+01  \\
& 3.904469026500e+01 & 6.885819592695e-02 & 8.899836120257e+01 & 3.681676812476e+00 & 1.694656422666e+02 & 2.881734287463e+02 & 3.28    \\
\hline
10& 4.174915157896e+01 & 1.922918402997e+01 & 6.960945806664e-02 & 9.165032030129e+01 & 9.061367433561e-03 & 2.933245693609e+02 & 2.875771933853e+01  \\
& 3.904495065324e+01 & 7.006912012210e-02 & 8.890436438806e+01 & 5.347656812186e+00 & 1.716994686435e+02 & 2.861362963230e+02 & 3.31    \\
\hline
11& 4.479530913553e+01 & 1.922882852351e+01 & 6.527265167414e-02 & 8.878969271572e+01 & 2.860240720175e-03 & 2.938135521314e+02 & 3.075854745471e+01  \\
& 3.904796816640e+01 & 7.425600491089e-02 & 9.004434245361e+01 & 7.072653584441e+00 & 1.685992005514e+02 & 2.896405828778e+02 & 3.36    \\
\hline
12& 4.154156511311e+01 & 1.922923085478e+01 & 6.896035181631e-02 & 9.117147245374e+01 & 4.071338029100e-03 & 2.933972196555e+02 & 2.861835343466e+01  \\
& 3.904464933657e+01 & 6.899287219873e-02 & 8.909957081781e+01 & 3.980319435622e+00 & 1.696051257475e+02 & 2.880512492534e+02 & 3.40    \\
\hline
13& 3.754325401968e+01 & 1.922938080353e+01 & 7.721696489983e-02 & 9.179579231219e+01 & 1.267974328728e+00 & 2.911834793866e+02 & 2.590420302446e+01  \\
& 3.904250288769e+01 & 6.569866758135e-02 & 8.892362876972e+01 & 4.841985920126e-01 & 1.681507155467e+02 & 2.890953991687e+02 & 3.52    \\
\hline
14& 4.159718920804e+01 & 1.922922312747e+01 & 6.882570653415e-02 & 9.100362137613e+01 & 2.470595788149e-04 & 2.934194355870e+02 & 2.866336688069e+01  \\
& 3.904473318052e+01 & 6.889161024838e-02 & 8.889847265918e+01 & 3.846188830491e+00 & 1.694417280429e+02 & 2.882055060155e+02 & 3.55    \\
\hline
15& 3.861928765858e+01 & 1.922895389064e+01 & 8.164332204177e-02 & 9.099047065022e+01 & 4.650033130131e+00 & 2.873274174741e+02 & 2.656823822493e+01  \\
& 3.904508142182e+01 & 7.876846440453e-02 & 9.065653510629e+01 & 1.011028992911e+01 & 1.733101135268e+02 & 2.855816625646e+02 & 4.07    \\
\hline
16& 3.665167058505e+01 & 1.922949066711e+01 & 7.954175382743e-02 & 9.176632661085e+01 & 1.984639542964e+00 & 2.902142061102e+02 & 2.528846331587e+01  \\
& 3.904165525712e+01 & 6.571095585591e-02 & 8.888129974765e+01 & 1.610534909587e+00 & 1.687342199023e+02 & 2.885166868435e+02 & 4.15    \\
\hline
17& 4.398306301341e+01 & 1.922888694344e+01 & 6.839621557165e-02 & 8.819762295219e+01 & 1.794595442165e+00 & 2.916646860913e+02 & 3.018841496602e+01  \\
& 3.904744109709e+01 & 7.631231392149e-02 & 9.046251959622e+01 & 8.337658507624e+00 & 1.676804661496e+02 & 2.907512204190e+02 & 4.31    \\
\hline
18& 4.497390073144e+01 & 1.922883287103e+01 & 6.479261194817e-02 & 8.954566028408e+01 & 4.530326498107e-03 & 2.938643925455e+02 & 3.089663494973e+01  \\
& 3.904791792625e+01 & 7.488168123117e-02 & 9.005183678814e+01 & 7.281033267811e+00 & 1.694884117770e+02 & 2.888473812254e+02 & 4.42    \\
\hline
19& 3.517320771351e+01 & 1.922927536404e+01 & 9.155836387497e-02 & 9.060354308371e+01 & 6.762090972125e+00 & 2.841505610261e+02 & 2.421977250287e+01  \\
& 3.904231833250e+01 & 7.704351101995e-02 & 9.032326871974e+01 & 1.009763604896e+01 & 1.729342455350e+02 & 2.857367320633e+02 & 4.44    \\
\hline
20& 3.612918368154e+01 & 1.922949360159e+01 & 8.197897082684e-02 & 9.179873009500e+01 & 2.683028114003e+00 & 2.892402802610e+02 & 2.493291020654e+01  \\
& 3.904143757622e+01 & 6.698518271539e-02 & 8.887479602283e+01 & 4.042533989682e+00 & 1.708061719417e+02 & 2.866549994566e+02 & 4.57    \\
\hline
\end{tabular}
\end{minipage}}
\label{tab3}
\end{table}

\begin{table}[htbp]
\centering
\rotatebox{90}{%
\begin{minipage}{0.95\textheight}
  \def\arraystretch{1.2}
   \small\centering
   \setlength{\tabcolsep}{1.0mm}
    \scriptsize
    \centering
\caption{The full precision parameter values for the second half of the 40 solutions in Table \ref{tab2}.}
\label{tab:k9_param}
\begin{tabular}{@{}cccccccc@{}}
\hline
Num. &$M_b$ ($M_\oplus$) &$P_b$ (days) & $e_b$ & $i_b$ & $\omega_b$ & $M_{0~ b}$ & $M_c$ ($M_\oplus$) \\
& $P_c$ (days) & $e_c$ & $i_c$ & $\Omega_b$-$\Omega_c$ & $\omega_c$ & $M_{0~ c}$ & $\chi^2$  \\
\hline
21& 3.530851724644e+01 & 1.922950484232e+01 & 8.697333887890e-02 & 9.145897667542e+01 & 4.791792749430e+00 & 2.866090229096e+02 & 2.435220537671e+01  \\
& 3.904114399052e+01 & 7.128509713375e-02 & 8.904777307171e+01 & 7.250415612891e+00 & 1.733064630843e+02 & 2.846971534201e+02 & 4.92    \\
\hline
22& 3.604528209006e+01 & 1.922923866587e+01 & 8.871963280967e-02 & 9.151441468979e+01 & 5.758243939940e+00 & 2.854643522907e+02 & 2.483618178510e+01  \\
& 3.904270703753e+01 & 7.652315932495e-02 & 8.907432299128e+01 & 9.859653339399e+00 & 1.762671081911e+02 & 2.824058076014e+02 & 5.46    \\
\hline
23& 3.515634391358e+01 & 1.922960746935e+01 & 8.353342093511e-02 & 9.178982571860e+01 & 2.896197964712e+00 & 2.888558626643e+02 & 2.426082340542e+01  \\
& 3.904061174052e+01 & 6.488662941556e-02 & 8.892553675051e+01 & 1.025051429714e-01 & 1.676260597338e+02 & 2.894994196806e+02 & 5.49    \\
\hline
24& 3.427152994230e+01 & 1.922942283987e+01 & 9.361881550724e-02 & 9.153476388464e+01 & 6.541865970919e+00 & 2.841495388896e+02 & 2.363287377936e+01  \\
& 3.904117816261e+01 & 7.473933152324e-02 & 8.903877779760e+01 & 9.464575054996e+00 & 1.757753077049e+02 & 2.826778878605e+02 & 5.87    \\
\hline
25& 3.625879321941e+01 & 1.922906109912e+01 & 9.078857177327e-02 & 9.084703888994e+01 & 6.820629141730e+00 & 2.841872455053e+02 & 2.496464688844e+01  \\
& 3.904367599363e+01 & 7.979983294459e-02 & 8.911722432846e+01 & 1.129472151289e+01 & 1.765588897794e+02 & 2.825076734009e+02 & 5.97    \\
\hline
26& 3.464626693196e+01 & 1.922933095302e+01 & 9.388255185139e-02 & 9.151749284138e+01 & 6.756000562351e+00 & 2.839179196469e+02 & 2.389236845540e+01  \\
& 3.904175976415e+01 & 7.631334219475e-02 & 8.904547562996e+01 & 1.020344873801e+01 & 1.765502539645e+02 & 2.821032553848e+02 & 6.13    \\
\hline
27& 3.619365538468e+01 & 1.922913036674e+01 & 9.074944054326e-02 & 9.130524147275e+01 & 6.579069137638e+00 & 2.844371105534e+02 & 2.492373379489e+01  \\
& 3.904332616691e+01 & 7.914821419564e-02 & 8.908482087345e+01 & 1.110978315146e+01 & 1.771258334742e+02 & 2.818695713870e+02 & 6.26    \\
\hline
28& 3.526762746214e+01 & 1.922923063988e+01 & 9.368722611548e-02 & 9.146303875538e+01 & 7.194035345565e+00 & 2.835094603447e+02 & 2.430527978968e+01  \\
& 3.904243531251e+01 & 7.866691485544e-02 & 8.903615783397e+01 & 1.110975427057e+01 & 1.772284286278e+02 & 2.817058879744e+02 & 6.48    \\
\hline
29& 4.577541728305e+01 & 1.922894984923e+01 & 6.340124403781e-02 & 8.964583305176e+01 & 1.411308685095e-03 & 2.940424736743e+02 & 3.143237216192e+01  \\
& 3.904778048011e+01 & 7.466457767238e-02 & 9.112665459198e+01 & 7.004757828961e+00 & 1.673861113239e+02 & 2.908480387098e+02 & 6.73    \\
\hline
30& 3.303388108739e+01 & 1.922975634741e+01 & 9.415736454380e-02 & 9.144806166199e+01 & 6.126231825863e+00 & 2.845044449623e+02 & 2.280796054340e+01  \\
& 3.903910575580e+01 & 6.957176934350e-02 & 8.892085118066e+01 & 7.076865671881e+00 & 1.728545787963e+02 & 2.849305141017e+02 & 6.83    \\
\hline
31& 3.347298745186e+01 & 1.922980421813e+01 & 8.863458610587e-02 & 9.172551399932e+01 & 4.164256579034e+00 & 2.870339904705e+02 & 2.311194542290e+01  \\
& 3.903905283551e+01 & 6.454179274547e-02 & 8.903235669298e+01 & 1.740055461626e+00 & 1.681203768502e+02 & 2.889688736930e+02 & 7.18    \\
\hline
32& 3.309145995884e+01 & 1.922987997689e+01 & 9.128933276625e-02 & 8.819607685851e+01 & 5.329931460347e+00 & 2.855975536669e+02 & 2.284723640400e+01  \\
& 3.903861485825e+01 & 6.589849645344e-02 & 9.112849146635e+01 & 3.779669424629e+00 & 1.653632300369e+02 & 2.917837808711e+02 & 7.47    \\
\hline
33& 3.282455083089e+01 & 1.922983192495e+01 & 9.119491163165e-02 & 9.178147056637e+01 & 4.620261983580e+00 & 2.862992510560e+02 & 2.266917952721e+01  \\
& 3.903867208743e+01 & 6.464263237635e-02 & 8.894576206224e+01 & 2.846777883133e+00 & 1.691263273760e+02 & 2.880043951577e+02 & 7.86    \\
\hline
34& 3.933281440991e+01 & 1.922915120751e+01 & 8.501385874946e-02 & 8.859154745458e+01 & 7.288584524863e+00 & 2.843736433988e+02 & 2.699717112720e+01  \\
& 3.904458276460e+01 & 8.248808461078e-02 & 9.087552297214e+01 & 1.221000187648e+01 & 1.693491751164e+02 & 2.898236735961e+02 & 8.17    \\
\hline
35& 4.711120232106e+01 & 1.922868952775e+01 & 6.244955058309e-02 & 8.944079636068e+01 & 4.237536308892e-03 & 2.941410843708e+02 & 3.229630947079e+01  \\
& 3.904951555731e+01 & 7.790150226611e-02 & 9.020701506683e+01 & 8.617930353200e+00 & 1.698098206011e+02 & 2.888776592265e+02 & 8.19    \\
\hline
36& 3.231349229781e+01 & 1.922997388983e+01 & 9.478686709237e-02 & 8.820212766892e+01 & 6.409897547552e+00 & 2.841507532913e+02 & 2.230092694368e+01  \\
& 3.903791318626e+01 & 6.682915400716e-02 & 9.108055087620e+01 & 4.927431468222e+00 & 1.650663013780e+02 & 2.921641882905e+02 & 8.23    \\
\hline
37& 4.222628404874e+01 & 1.922867666628e+01 & 7.593626590446e-02 & 9.098864200808e+01 & 3.641164918238e+00 & 2.889890821889e+02 & 2.901213009875e+01  \\
& 3.904768852342e+01 & 8.220413052828e-02 & 8.943657151849e+01 & 1.149292939731e+01 & 1.763048587544e+02 & 2.830480241397e+02 & 8.39    \\
\hline
38& 3.209867768901e+01 & 1.922995986346e+01 & 9.557741304010e-02 & 8.821352911998e+01 & 6.521452656086e+00 & 2.839506090679e+02 & 2.215676936115e+01  \\
& 3.903788467827e+01 & 6.680281691352e-02 & 9.112766716924e+01 & 4.973505595475e+00 & 1.650736333005e+02 & 2.921591442350e+02 & 8.51    \\
\hline
39& 3.217568388089e+01 & 1.923001392814e+01 & 9.495597001914e-02 & 8.819871437657e+01 & 6.301489337750e+00 & 2.842455230402e+02 & 2.221965214291e+01  \\
& 3.903765426852e+01 & 6.618452780251e-02 & 9.111987201307e+01 & 4.529629715965e+00 & 1.649372420951e+02 & 2.922137658765e+02 & 8.58    \\
\hline
40& 3.158324632275e+01 & 1.923000403900e+01 & 9.765501951550e-02 & 8.820089509382e+01 & 6.675574937299e+00 & 2.835751385343e+02 & 2.180392302664e+01  \\
& 3.903750468964e+01 & 6.628218832921e-02 & 9.110782075731e+01 & 5.064158700987e+00 & 1.651091919348e+02 & 2.920599502435e+02 & 9.08    \\
\hline
\end{tabular}
\end{minipage}}
\label{tab3}
\end{table}

\end{document}